\documentclass[conference]{IEEEtran}

\usepackage{amsmath,amssymb,epsfig,psfrag,cite,subfigure}
\include{macros}
\usepackage{graphicx}
\usepackage{epstopdf}

\usepackage{acronym}

\usepackage{bm}

\usepackage{pifont}

\usepackage{tikz}
\usetikzlibrary{calc}

\usepackage{tcolorbox}

\usepackage{accents}

\allowdisplaybreaks

\usepackage{soul}

\usepackage{amsthm}

\usepackage{nicefrac}

\usepackage{lipsum,multicol}

\usepackage{algorithm}
\usepackage{algpseudocode}

\algdef{SE}[SUBALG]{Indent}{EndIndent}{}{\algorithmicend\ }%
\algtext*{Indent}
\algtext*{EndIndent}

\usepackage{enumitem}
\usepackage{mwe}    
\usepackage{epsfig,psfrag}
\usepackage{subfigure}
\usepackage{color}
\usepackage{url}
\usepackage{mathtools,xparse}

\usepackage{gensymb}

\usepackage[multiple]{footmisc}

\usepackage{xr}

\makeatletter
\newcommand*{\addFileDependency}[1]{
  \typeout{(#1)}
  \@addtofilelist{#1}
  \IfFileExists{#1}{}{\typeout{No file #1.}}
}
\makeatother

\newcommand*{\myexternaldocument}[1]{%
    \externaldocument{#1}%
    \addFileDependency{#1.tex}%
    \addFileDependency{#1.aux}%
}

\myexternaldocument{supp}

\usepackage{xcolor}

\newcommand\abs[1]{\big|#1\big|}
\newcommand\abss[1]{\lvert#1\rvert}
\newcommand\norm[1]{\left\lVert #1\right\rVert}

\newcommand{\thn}[1]{ {#1^{\rm{th} } } }

\newcommand{\Eee}{\mathbb{E}}

\newcommand{\snrx}{{\rm{SNR}}_{\rm{x}}}
\newcommand{\snrh}{{\rm{SNR}}_{\rm{h}}}

\newcommand{\pfa}{P_{\rm{fa}}}
\newcommand{\pd}{P_{\rm{d}}}

\newcommand{\Tsym}{ T_{\rm{sym}} }

\newcommand{\Tcp}{ T_{\rm{cp}} }

\newcommand{\FF}{ \mathbf{F} }

\newcommand{\deltaf}{ \Delta f }

\newcommand{\fc}{ f_c }
\newcommand{\Ntx}{ N_{\rm{T}} }

\newcommand{\atx}{ \aaa_{\rm{T}} }

\newcommand{\HHhat}{ \widehat{\HH} }

\newcommand{\xnm}{ x_{n,m} }

\newcommand{\znm}{ z_{n,m} }

\newcommand{\ptx}{\pp_{\rm{T}}}
\newcommand{\prx}{\pp_{\rm{R}}}
\newcommand{\ptk}{\pp_k}

\newcommand{\boldHhat}{ \widehat{\boldH} }

\newcommand{\fftx}{ \ff_{\rm{T}} }

\newcommand{\alphahat}{ \widehat{\alpha} }

\newcommand{\alphahatb}{ \boldsymbol{\alphahat}}

\newcommand{\Iind}{ \mathcal{I} }

\newcommand{\mtCN}{{\mathcal{CN}}}

\newcommand{\mP}{ \mathcal{P} }
\newcommand{\mD}{ \mathcal{D} }
\newcommand{\mG}{ \mathcal{G} }

\newcommand{\prat}{\rho_{\mP}}
\newcommand{\XXp}{\XX_{\mP}}
\newcommand{\XXd}{\XX_{\mD}}

\newcommand{\veccs}[1]{ {\rm{vec}}\big(#1\big)  }

\newcommand{\aaa}{\mathbf{a}}
\newcommand{\cc}{ \mathbf{c} }
\newcommand{\bb}{ \mathbf{b} }

\newcommand{\nuhat}{{ \widehat{\nu} }}
\newcommand{\tauhat}{{ \widehat{\tau} }}

\newcommand{\boldzero}{{ {\boldsymbol{0}} }}

\newcommand{\Pt}{{ P_{\rm{T}} }}

\newcommand{\sigmarcsk}{{ \sigma_{{\rm{rcs}},k} }}

\newcommand{\ff}{\mathbf{f}}

\newcommand{\YY}{ \mathbf{Y} }

\newcommand{\HH}{ \mathbf{H} }

\newcommand{\hhatdd}{ \boldHhat^{\rm{DD}} }

\newcommand{\hhatdds}{ \widehat{h}^{\rm{DD}} }

\newcommand{\XX}{ \mathbf{X} }

\newcommand{\ZZ}{ \mathbf{Z} }

\newcommand{\boldH}{ \mathbf{H} }

\newcommand{\boldA}{ \mathbf{A} }

\newcommand{\pp}{ \mathbf{p} }

\newcommand{\hh}{ \mathbf{h} }

\newcommand{\vv}{ \mathbf{v} }

\newcommand{\trp}{\mathsf{T}}
\newcommand{\herm}{\mathsf{H}}

\newcommand{\cset}[2]{ \mathbb{C}^{#1 \times #2}  }
\newcommand{\csett}{ \mathbb{C}  }
\newcommand{\rset}[2]{ \mathbb{R}^{#1 \times #2}  }

\newcommand{\conj}{ {\ast} }

\newcommand{\Khat}{\widehat{K}}

\newcommand{\XXdhat}{ \widehat{\mathbf{X}}_{\mD} }

\newcommand{\YYdhat}{ \widehat{\mathbf{Y}}_{\mD} }
\newcommand{\HHdhat}{ \widehat{\mathbf{H}}_{\mD} }
\newcommand{\HHphat}{ \widehat{\mathbf{H}}_{\mP} }

\makeatletter \renewcommand\d[1]{\ensuremath{%
		\;\mathrm{d}#1\@ifnextchar\d{\!}{}}}
\makeatother

\makeatletter
\newcommand*\rel@kern[1]{\kern#1\dimexpr\macc@kerna}
\newcommand*\widebar[1]{%
  \begingroup
  \def\mathaccent##1##2{%
    \rel@kern{0.8}%
    \overline{\rel@kern{-0.8}\macc@nucleus\rel@kern{0.2}}%
    \rel@kern{-0.2}%
  }%
  \macc@depth\@ne
  \let\math@bgroup\@empty \let\math@egroup\macc@set@skewchar
  \mathsurround\z@ \frozen@everymath{\mathgroup\macc@group\relax}%
  \macc@set@skewchar\relax
  \let\mathaccentV\macc@nested@a
  \macc@nested@a\relax111{#1}%
  \endgroup
}
\makeatother

\theoremstyle{remark}

\newtheoremstyle{mytheoremstyle} 
    {\topsep}                    
    {\topsep}                    
    {\upshape}                   
    {.5em}                           
    {\itshape}                   
    {.}                          
    {.5em}                       
    {}  

\theoremstyle{plain}

\newtheoremstyle{iremark}
  {\topsep}   
  {\topsep}   
  {\upshape}  
  {0.2in}       
  {\itshape}  
  {.}         
  {5pt plus 1pt minus 1pt} 
  {\thmname{#1}\thmnumber{ \itshape#2}\thmnote{ (#3)}} 

\theoremstyle{definition}
\newtheorem*{proof}{Proof}

\theoremstyle{definition}

\acrodef{RIS}{reconfigurable intelligent surface}
\acrodef{RCS}{radar cross section}
\acrodef{SNR}{signal-to-noise ratio}
\acrodef{ISAC}{integrated sensing and communication}
\acrodef{ISLAC}{integrated sensing, localization, and communication}
\acrodef{LOS}{line-of-sight}
\acrodef{NLOS}{non-line-of-sight}
\acrodef{AOA}{angle-of-arrival}
\acrodef{AOD}{angle-of-departure}
\acrodef{UE}{user equipment}
\acrodef{NF}{near-field}
\acrodef{BS}{base station}
\acrodef{MCRB}{misspecified Cram\'{e}r-Rao bound}
\acrodef{CRB}{Cram\'{e}r-Rao bound}
\acrodef{LB}{lower bound}
\acrodef{ML}{maximum-likelihood}
\acrodef{MML}{mismatched maximum-likelihood}
\acrodef{DL}{downlink}
\acrodef{UL}{uplink}
\acrodef{MIMO}{multiple-input multiple-output}
\acrodef{MISO}{multiple-input single-output}
\acrodef{SISO}{single-input single-output}
\acrodef{SIP}{shift invariance property}
\acrodef{FIM}{Fisher information matrix}
\acrodef{RMSE}{root mean-squared error}
\acrodef{AWGN}{additive white Gaussian noise}
\acrodef{ADMM}{alternating direction method of multipliers}
\acrodef{LS}{least-squares}
\acrodef{SOC}{second-order cone}
\acrodef{DoF}{degree-of-freedom}
\acrodef{CFO}{carrier frequency offset}
\acrodef{TO}{timing offset}
\acrodef{ICI}{intercarrier interference}
\acrodef{i.i.d.}{independently and identically distributed}
\acrodef{MI}{mutual information}
\acrodef{SAC}[S\&C]{sensing and communication}

\acrodef{ULA}{uniform linear array}

\acrodef{RF}{reciprocal filtering}
\acrodef{MF}{matched filtering}
\acrodef{LMMSE}{linear minimum mean-squared-error}

\acrodef{SSB}{synchronization signal block}
\acrodef{DMRS}{demodulation reference signal}
\acrodef{PTRS}{phase tracking reference signal}

\acrodef{LO}{local oscillator}
\acrodef{PPM}{parts-per-million} 

\graphicspath{{./Figures/}}

\IEEEoverridecommandlockouts

\begin{document}
\bstctlcite{IEEEexample:BSTcontrol}

\title{Bridging the Gap via Data-Aided Sensing: Can Bistatic ISAC Converge to Genie Performance?}

\author{
\IEEEauthorblockN{
Musa Furkan Keskin\IEEEauthorrefmark{1}\thanks{This work is partially supported by the Vinnova B5GPOS project under Grant 2022-01640, the SNS JU project 6G-DISAC under the EU’s Horizon Europe research and innovation programme under Grant Agreement No 101139130, the Swedish Research Council (VR) through the project 6G-PERCEF under Grant 2024-04390, and by the European Union under the Italian National
Recovery and Resilience Plan (NRRP) of NextGenerationEU, partnership on
“Telecommunications of the Future” (PE00000001 - program “RESTART”).},
Silvia Mura\IEEEauthorrefmark{2},
Marouan Mizmizi\IEEEauthorrefmark{2},
Dario Tagliaferri\IEEEauthorrefmark{2},
Henk Wymeersch\IEEEauthorrefmark{1}
}
\IEEEauthorblockA{
\IEEEauthorrefmark{1}Department of Electrical Engineering, Chalmers University of Technology, Sweden\\
\IEEEauthorrefmark{2}Department of Electronics, Information and Bioengineering, Politecnico di Milano, Italy
}
}


\maketitle


\begin{abstract}
We investigate data-aided iterative sensing in bistatic OFDM ISAC systems, focusing on scenarios with co-located sensing and communication receivers. To enhance target detection beyond pilot-only sensing methods, we propose a multi-stage bistatic OFDM receiver, performing iterative sensing and data demodulation to progressively refine ISAC channel and data estimates. Simulation results demonstrate that the proposed data-aided scheme significantly outperforms pilot-only benchmarks, particularly in multi-target scenarios, substantially narrowing the performance gap compared to a genie-aided system with perfect data knowledge. Moreover, the proposed approach considerably expands the bistatic ISAC trade-off region, closely approaching the probability of detection–achievable rate boundary established by its genie-aided counterpart.

	\textit{Index Terms--} OFDM, ISAC, bistatic ISAC, bistatic sensing, data-aided sensing.
\end{abstract}

\section{Introduction}

Integrated sensing and communication (ISAC) has emerged as a key enabler for next-generation (6G) wireless networks, enabling simultaneous data transmission and environmental sensing \cite{Fan_ISAC_6G_JSAC_2022}. Among ISAC configurations, \textit{monostatic ISAC}, where the transmitter and receiver are co-located, has been extensively studied due to inherent advantages, such as the use of communication data for sensing and the absence of synchronization requirements \cite{5G_NR_JRC_analysis_JSAC_2022, ISAC_OFDM_Pilots,Liu_Reshaping_OFDM_ISAC_2023}. On the other hand, \textit{bistatic ISAC}, involving a spatially separated transmitter and receiver, has recently gained traction \cite{bistatic_OFDM_ISAC_OTA_2024,bistaticISAC_Clock_Async_2024,bistatic_ISAC_joint_2024} as it offers specific benefits over monostatic ISAC, including elimination of full-duplex transceiver complexity, extended sensing coverage and enhanced target detection through spatial diversity \cite{bistatic_ISAC_Andrea_2022,bistaticISAC_Clock_Async_2024}.

The spatially separated architecture of bistatic ISAC introduces several technical challenges, such as stringent synchronization requirements between devices and the need for efficient data/pilot resource allocation \cite{bistatic_ISAC_exp_2023,bistaticOFDM_5G_NR_TVT_2024}. While resource-allocation strategies balancing pilot and data usage have been investigated in joint localization and communication networks \cite{ISACNet}, these methods primarily focus on cooperative scenarios, where users/targets actively participate in the communication process, inherently avoiding issues such as multi-target interference \cite{holisticISAC_2024}. In contrast, bistatic ISAC often involves non-cooperative passive targets, such as unknown objects that do not emit signals, rendering conventional cooperative methods inadequate \cite{bistatic_OFDM_ISAC_OTA_2024}. Another key issue with bistatic ISAC is that the transmit signal is only partially known at the receiver since typically only the training pilots used for demodulation purposes are known, while the data symbols remain unknown \cite[Sec.~VI-B2c]{Fan_ISAC_6G_JSAC_2022}. Relying solely on these sparse pilots for sensing may introduce sidelobes in the sensing ambiguity function, which might degrade sensing performance, especially in multi-target scenarios \cite{ISAC_OFDM_Pilots}.

\begin{figure}[!t]
	\centering
	\includegraphics[width=0.7\linewidth]{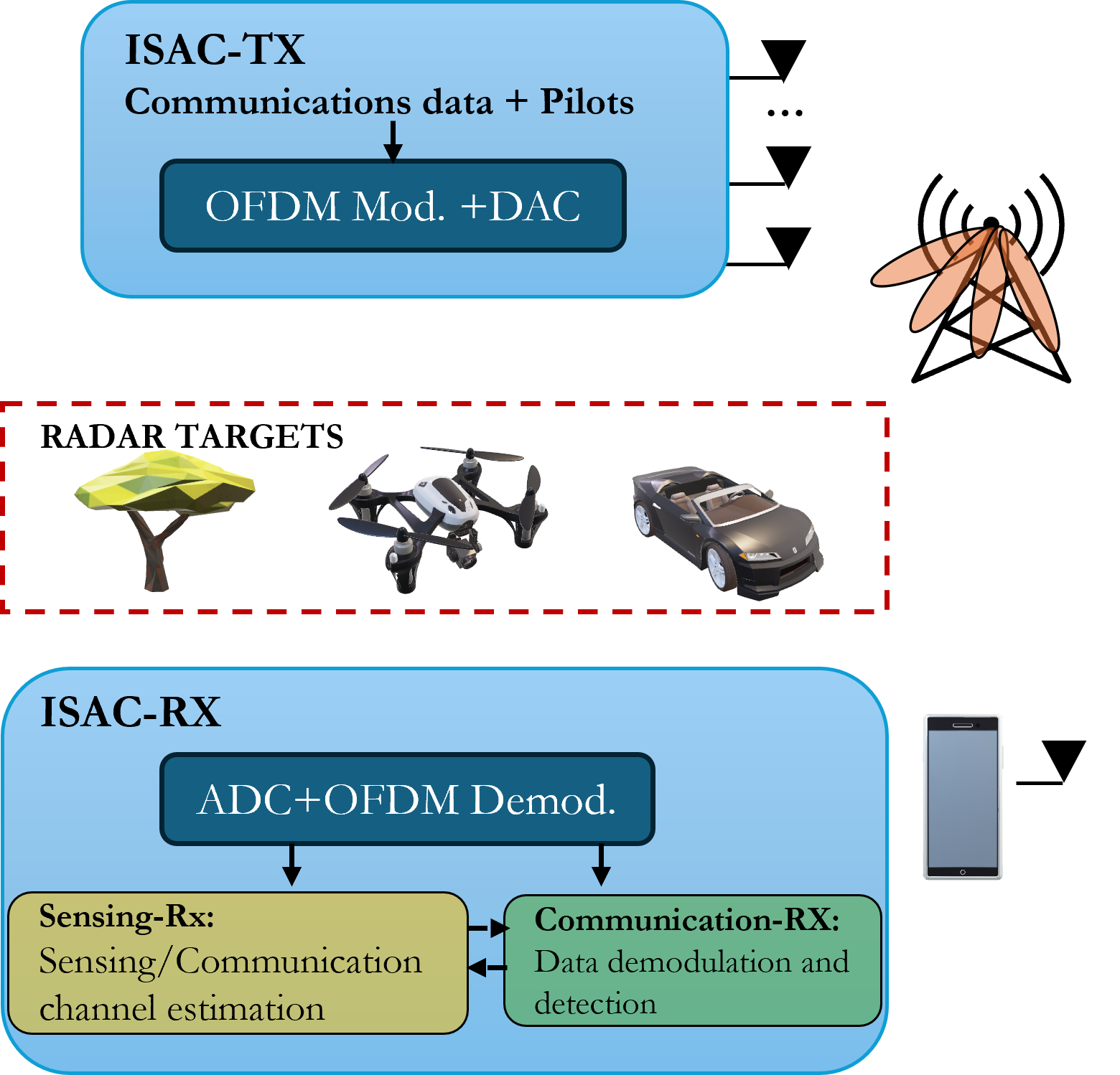}
	\caption{Bistatic ISAC scenario with co-located sensing and communication receivers at the ISAC receiver.} 
	\label{fig_scenario}
	\vspace{-0.1in}
\end{figure}

To tackle the issue of unknown data payload in bistatic ISAC, data-aided iterative channel estimation techniques have been proposed for \textit{single-carrier} \cite{bistatic_ISAC_joint_2022,bistatic_ISAC_joint_2024} and \textit{ultra wideband (UWB)} \cite{bistaticISAC_UWB_Liu_2024} systems, employing both pilot symbols and detected data to refine estimates of target parameters. The \textit{orthogonal frequency-division multiplexing (OFDM)} waveform, on the other hand, has received relatively little attention in data-aided bistatic ISAC investigations (e.g., \cite{bistatic_OFDM_ISAC_OTA_2024}) despite its wide adoption in 5G and future 6G deployments \cite{SPM_ISAC_multicarrier_2024}. In \cite{bistatic_OFDM_ISAC_OTA_2024}, the impact of bit errors on delay-Doppler images has been qualitatively analyzed without explicitly evaluating crucial ISAC metrics, such as detection performance and achievable data rate. Consequently, several fundamental aspects remain unexplored in data-aided bistatic OFDM ISAC systems: \textit{(i)} multi-target detection performance and the associated target masking effects \cite{Liu_Reshaping_OFDM_ISAC_2023}, \textit{(ii)} impact of modulation order for data on sensing performance \cite{holisticISAC_2024}, \textit{(iii)} how pilot allocation affects both probability of detection  and data rate, and \textit{(iv)} a systematic investigation of bistatic ISAC trade-offs.

Motivated by these research gaps, this paper proposes a data-aided sensing framework for bistatic OFDM ISAC systems, systematically analyzing the trade-offs between radar target detection performance and communication rate. Our goal is to answer a fundamental question: \textit{to what extent and under which conditions can the performance of a bistatic OFDM ISAC system approach that of its genie-aided counterpart with perfect knowledge of data symbols?} Our contributions are as follows:
\begin{itemize}
    \item We design a multi-stage, iterative data-aided receiver for bistatic OFDM ISAC systems, enabling simultaneous and progressively refined radar sensing and data demodulation.
    \item We analyze the data-aided multi-target detection performance under various conditions, including target \ac{RCS}, pilot percentage and choice of OFDM ISAC channel estimator (i.e., \ac{RF}, \ac{MF} or \ac{LMMSE} \cite{OFDM_Radar_Corr_TAES_2020,holisticISAC_2024}), benchmarking against both pilot-only and genie-aided schemes.  
    \item We investigate bistatic ISAC trade-offs between probability of detection and achievable rate across different modulation schemes (QPSK vs. high-order QAM), highlighting the extent to which data-aided sensing can bridge the performance gap between pilot-only and genie-aided schemes.   
\end{itemize}

\section{System Model and Problem Formulation}
In this section, we provide the scenario description, signal model and problem formulation for the bistatic ISAC setup.

\subsection{Scenario Description}
We consider a bistatic ISAC scenario consisting of two entities, namely \textit{(i)} a multiple-antenna ISAC transmitter (ISAC-TX) and \textit{(ii)} a single-antenna ISAC receiver (ISAC-RX), as well as a number of targets in the environment, as shown in Fig~\ref{fig_scenario}. ISAC-TX communicates with ISAC-RX by transmitting both data and pilots within a given OFDM time-frequency resource. In the considered bistatic configuration, ISAC-RX contains a sensing receiver (S-RX) and a communication receiver (C-RX) co-located on the same platform, and performs bistatic sensing and communication receive operations simultaneously \cite{bistaticISAC_UWB_Liu_2024,bistatic_ISAC_joint_2024}. Given the unified received signal at ISAC-RX, the goal of S-RX is to detect the targets in the environment and estimate their delay and Doppler parameters, while C-RX carries out OFDM data demodulation and data detection tasks. Sharing the same device, S-RX and C-RX mutually benefit each other to enhance bistatic sensing and communication performances via information exchange and joint optimization (i.e., \textit{device-level mutual assistance} \cite{Fan_ISAC_6G_JSAC_2022}).

\subsection{ISAC Signal Model}
For the purpose of joint \ac{DL} communications with ISAC-RX and bistatic sensing, ISAC-TX employs an OFDM waveform with $N$ subcarriers and $M$ symbols. The total duration of each symbol is $\Tsym = \Tcp + T$, with $\Tcp$, $T$ and $\deltaf = 1/T$ denoting the cyclic prefix (CP) duration, the elementary symbol duration and the subcarrier spacing, respectively \cite{RadCom_Proc_IEEE_2011}. Since sensing and communication receivers are co-located at ISAC-RX, the bistatic ISAC scenario under consideration features a \textit{unified received signal model} for sensing and communications \cite{bistatic_ISAC_joint_2022}. The baseband received signal at ISAC-RX on subcarrier $n$ and symbol $m$ is given by \cite{RadCom_Proc_IEEE_2011,procIEEE_2024_Nuria}
\begin{align} \label{eq_ynm}
    & y_{n,m} = \hh^\trp_{n,m} \fftx \xnm + \znm   \,,
\end{align}
where $\fftx \in \cset{\Ntx}{1}$ is the beamforming vector at the ISAC-TX array\footnote{For the sake of simplicity, we employ a fixed beamformer $\fftx$ over the entire frame of $M$ symbols. Studying the impact of time-varying beamforming design on ISAC performance is left as a future work.}, with $\Ntx$ denoting the number of antenna elements, $\xnm \in \csett$ is the data/pilot on subcarrier $n$ and symbol $m$, $\znm \sim \mtCN(0, \sigma^2) $ denotes the additive white Gaussian noise (AWGN) and $\sigma^2 = N_0 N \deltaf N_F$ with $N_0$ and $N_F$ representing the noise power spectral density (PSD) and the noise figure, respectively, and $\hh_{n,m} \in \cset{\Ntx}{1}$ is the unified ISAC channel on subcarrier $n$ and symbol $m$ \cite{LiuSeventy23,SPM_ISAC_multicarrier_2024}:
\begin{align} \label{eq_hh_vec}
    \hh_{n,m} = \sum_{k=0}^{K}  \alpha_k e^{-j 2 \pi n \deltaf \tau_{k} }  e^{j 2 \pi  m \Tsym \nu_{k}} \atx(\theta_k) \,. 
\end{align}

In \eqref{eq_ynm}, we set $\norm{\fftx}^2 = \Pt$ and $\Eee\{ \abss{\xnm}^2 \} = 1$, where $\Pt$ denotes the transmit power. As to \eqref{eq_hh_vec}, we consider the presence of $K$ targets/scatterers in the ISAC channel between ISAC-TX and ISAC-RX, along with a \ac{LOS} path. The $\thn{k}$ path is characterized by a complex channel gain $\alpha_k$ (including the effects of bistatic radar cross section for $k>0$ and path attenuation), an \ac{AOD} $\theta_k$, a delay $\tau_k$ and a Doppler shift $\nu_k$, where $k=0$ denotes the \ac{LOS} path while $k>0$ corresponds to paths induced by the scattering of the ISAC signal off the targets.\footnote{Here, $\tau_k$ and $\nu_k$ represent the \textit{ambiguous} delay and Doppler shift of the $\thn{k}$ path, respectively, which differ from their \textit{true} values by the amount of \ac{TO} and \ac{CFO} due to clock asynchronism  between ISAC-TX and ISAC-RX \cite{bistaticISAC_Clock_Async_2024}. Our focus will be on estimating $\tau_k$ and $\nu_k$, while the task of estimating the position and velocity of the scatterers as well as the \ac{TO} and \ac{CFO} based on the estimates of $\tau_k$ and $\nu_k$ is deferred to a subsequent stage (outside the scope of this work) \cite{grossi2020adaptive}.} 
Moreover, $\atx(\theta) \in \cset{\Ntx}{1}$ denotes the steering vector of the \ac{ULA} at ISAC-TX with $[\atx(\theta)]_i = e^{j \frac{2 \pi}{\lambda} i d \sin(\theta)}$, $\lambda = c/\fc$, $c$ and $d$ denote the wavelength, speed of propagation and element spacing, respectively. Additionally, in obtaining \eqref{eq_ynm},  we leverage the following standard assumptions: \textit{(i)} the delay spread of the ISAC channel is smaller than the CP\footnote{Considering a standard 5G NR numerology with $\deltaf = 60 \, \rm{kHz}$ \cite[Sec.~4.2]{TR_38211}, the CP is calculated as $\Tcp = 0.07/\deltaf = 1.16 \, \rm{\mu s}$. This corresponds to a maximum distance spread of $350 \, \rm{m}$ in bistatic sensing, which can accommodate a broad spectrum of channel conditions encountered in 5G/6G communication scenarios \cite{SPM_ISAC_multicarrier_2024}.}, i.e., $\Tcp \geq \max_k \tau_k - \min_k \tau_k$  \cite{SPM_JRC_2019,SPM_ISAC_multicarrier_2024}, \textit{(ii)} $\max_{k} \abss{\nu_{k}} \ll \fc/(NM)$ \cite{MIMO_OFDM_ICI_JSTSP_2021,ICI_OFDM_TSP_2020} (the narrowband approximation), and \textit{(iii)} the fast-time phase rotations within a symbol are negligible \cite{bistaticISAC_Clock_Async_2024,jump_imdea_2023}.

Stacking \eqref{eq_ynm} over $N$ subcarriers and $M$ symbols, and absorbing the TX beamforming gain $\atx^\trp(\theta_k) \fftx$ into $\alpha_k$ in \eqref{eq_hh_vec}, we obtain the frequency/slow-time observation matrix
\begin{align} \label{eq_YY}
    \YY &= \XX \odot \HH + \ZZ \in \cset{N}{M} \,,
\end{align}
where $\odot$ denotes Hadamard (element-wise) multiplication, $\YY \in \cset{N}{M}$ with $[\YY]_{n,m} \triangleq y_{n,m}$, $\XX \in \cset{N}{M}$ with $[\XX]_{n,m} \triangleq x_{n,m}$, $\ZZ \in \cset{N}{M}$ with $[\ZZ]_{n,m} \triangleq z_{n,m}$, and
\begin{align} \label{eq_hh}
    \HH &\triangleq \sum_{k=0}^{K}   \alpha_{k} \,  \bb(\tau_{k}) \cc^\trp(\nu_{k}) \in \cset{N}{M} \,.
\end{align}
Here, $\bb(\tau) \in \cset{N}{1}$ with $[\bb(\tau)]_n = e^{-j 2 \pi n \deltaf  \tau}$ and $\cc(\nu) \in \cset{M}{1}$ with  $[\cc(\nu)]_m = e^{j 2 \pi m \Tsym \nu }$ representing, respectively, the frequency-domain and time-domain steering vectors.

\subsection{Degrees-of-Freedom in Bistatic ISAC}
Given the unified sensing/communication signal model \eqref{eq_YY} at the bistatic ISAC receiver, we have two different degrees-of-freedom (DoFs) to optimize ISAC trade-offs.

\subsubsection{Data/Pilot Selection}
Let $\mP \subseteq \mG = \{(n,m) : 1 \leq n \leq N,\, 1 \leq m \leq M\}$ and $\mD = \mG \setminus \mP$ denote the set of subcarrier-symbol index pairs that indicate, respectively, pilot and data locations in the OFDM frequency-time grid $\mG$. Sensing and communication performances will be affected by both the percentage of pilots (e.g., effective data rate vs. radar SNR \cite{bistatic_OFDM_ISAC_OTA_2024}), defined as $\prat = 100 \, \abss{\mP}/(NM)$, and the placement of pilots (e.g., channel interpolation from an undersampled frequency-time grid \cite{ISAC_OFDM_Pilots} using random or periodic pilot pattern). For convenience, we define $\XX_{\mP}$ and $\XX_{\mD}$ as the submatrices of $\XX$ in \eqref{eq_YY} consisting of elements residing in the pilot and data locations, respectively.

\subsubsection{Modulation Order Selection}
In our setup, `pilot' denotes dedicated sensing pilots with unit amplitude \cite{dedicatedSensingISAC_2024}, while `data' may refer to either unit-amplitude (e.g., QPSK) or varying-amplitude (e.g., QAM) symbols. In view of the time-frequency trade-off in OFDM ISAC systems \cite{holisticISAC_2024,Liu_Reshaping_OFDM_ISAC_2023}, the modulation order selection for $\XXd$ constitutes another DoF to tune ISAC trade-offs in the considered bistatic scenario.

\subsection{Problem Statement for Bistatic ISAC}\label{sec_prob}
Given the received signal $\YY$ in \eqref{eq_YY} and the knowledge of pilot symbols $\XXp$, the problems of interest are \textit{(i)} to detect the presence of multiple targets/paths and estimate their parameters\footnote{Due to the use of a fixed beamformer $\fftx$ over the entire frame, estimation of \acp{AOD} $\{\theta_k\}_{k=0}^K$ is not possible.} $\{\alpha_k, \tau_k, \nu_k\}_{k=0}^K$ in \eqref{eq_hh}, and \textit{(ii)} perform data demodulation to estimate the unknown data symbols $\XXd$.

\section{Receiver Design in Bistatic ISAC}\label{sec_rec_design}
\begin{figure}
	\centering
    \vspace{-0.2in}
	\includegraphics[width=1\linewidth]{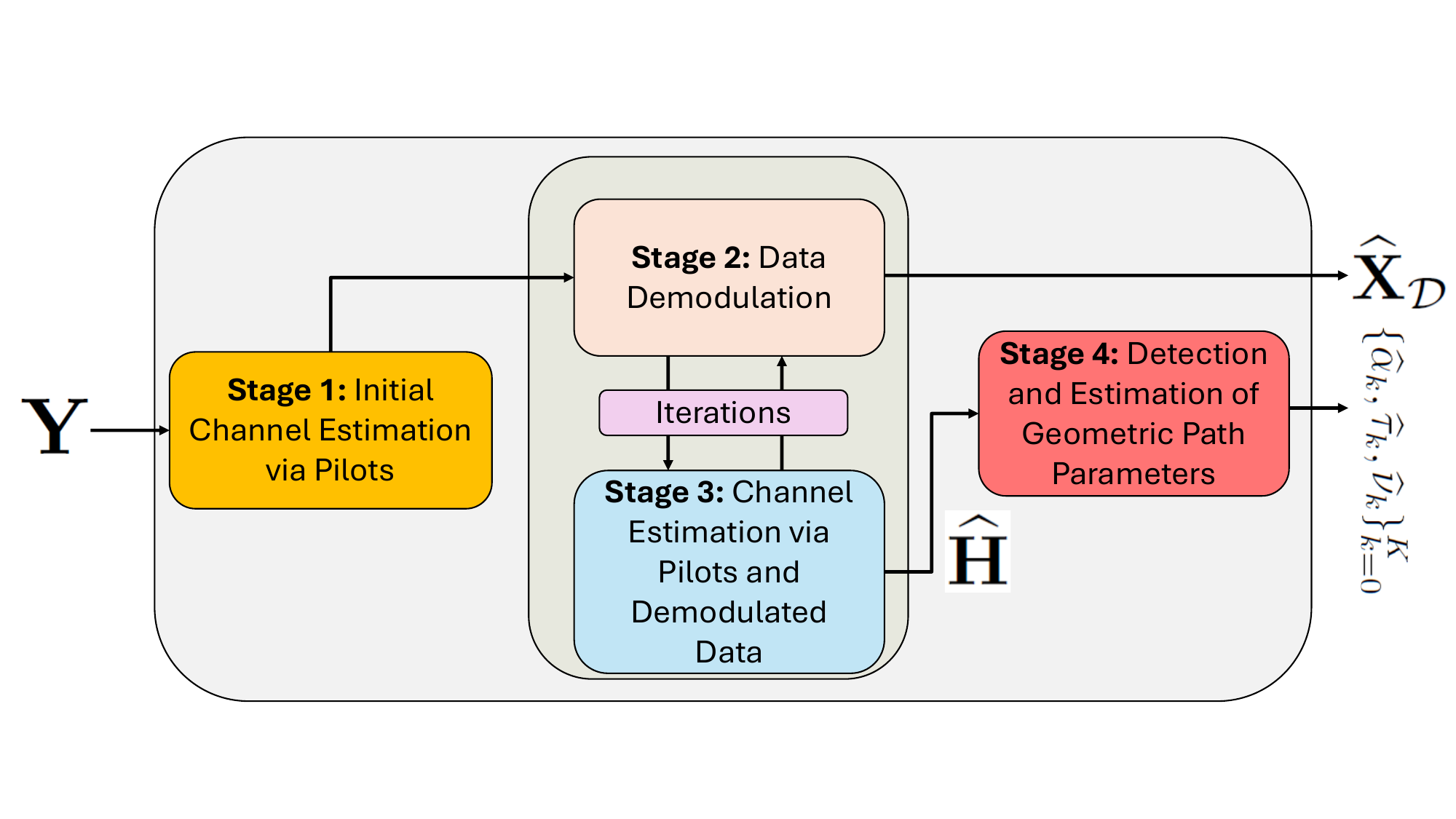}
	\vspace{-0.5in}
	\caption{The proposed multi-stage bistatic OFDM ISAC receiver that iteratively performs demodulation and data-aided sensing.} 
	\label{fig_algorithm}
	\vspace{-0.1in}
\end{figure}

In this section, we propose a multi-stage algorithm for the bistatic OFDM ISAC receiver to tackle the problem of joint target sensing and communications formulated in Sec.~\ref{sec_prob}. A high-level overview of the algorithm is given in Fig.~\ref{fig_algorithm}. We begin by estimating the unstructured ISAC channel matrix $\HH$ in \eqref{eq_YY} using only pilot symbols $\XXp$. In Stage~2, data demodulation is performed to obtain an estimate $\XXdhat$ of $\XXd$ using the estimate $\HHhat$ of $\HH$ from the first stage. Then, Stage~3 involves refining the channel estimate $\HHhat$ using both pilots $\XXp$ and demodulated data $\XXdhat$. Stage~2 and 3 are executed iteratively until convergence. In the final stage, we perform target detection and parameter estimation from $\HHhat$ by exploiting its inner geometric structure in \eqref{eq_hh}. In the following, we elaborate on the different stages of the proposed scheme. 

\subsection{Stage 1: Initial Channel Estimation via Pilots}\label{sec_sub1_initial}
Following a similar approach to \cite[Sec.~VI]{ISAC_OFDM_Pilots}, we design a pilot-only channel estimation strategy for Stage~1 that can be decomposed into several substages.
\subsubsection{Channel Estimation at Pilot Locations}\label{sec_ce_pilot1}
We construct an estimate $\HHhat$ from \eqref{eq_YY} by estimating the channel at pilot locations via standard \ac{RF} and setting the data locations to $0$:
\begin{subequations} \label{eq_hh_s11}
\begin{align} \label{eq_hhhat_p}
    \HHhat_{\mP} &= \YY_{\mP} \oslash \XXp \, ,
    \\
    \HHhat_{\mD} &= [\boldzero_{N\times M}]_{\mD} \,,
\end{align}
\end{subequations}
where $\oslash$ denotes element-wise division.

\subsubsection{Target Detection/Estimation from Channel Estimate}\label{sec_det_est}
Using the structure in \eqref{eq_hh}, we form the delay-Doppler image from $\HHhat$ in \eqref{eq_hh_s11} \cite{RadCom_Proc_IEEE_2011,MIMO_OFDM_ICI_JSTSP_2021}
\begin{align}
    \hhatdds(\tau, \nu) &= \abs{ \bb^\herm(\tau) \HHhat \cc^\conj(\nu) }^2 \,, 
\end{align}
which can be implemented via 2-D FFT \cite{ISAC_OFDM_Pilots,holisticISAC_2024}:
\begin{align} \label{eq_hhatdd}
    \hhatdd &= \abs{  \FF_N^\herm \HHhat \FF_M   }^2 \in \rset{N}{M} \,,
\end{align}
where $\FF_N \in \cset{N}{N}$ is the unitary DFT matrix. Detection of multiple targets and estimation of their delay-Doppler parameters can be achieved by identifying peaks in \eqref{eq_hhatdd} using a constant false alarm rate (CFAR) detector, as described in \cite[Ch.~6.2.4]{richards2005fundamentals}, yielding $\{\tauhat_k, \nuhat_k\}_{k=0}^{\Khat}$.

\subsubsection{Channel Reconstruction from Target Estimates}
The gain estimates corresponding to $\{\tauhat_k, \nuhat_k\}_{k=0}^{\Khat}$ can be obtained via \ac{LS} using $\HHhat$ in \eqref{eq_hh_s11} and the structure in \eqref{eq_hh} \cite{holisticISAC_2024}:
\begin{align}
    \alphahatb = \boldA^{\dagger} \veccs{\HHhat} \in \cset{(\Khat+1)}{1} \,,
\end{align}
where $\boldA = [\aaa_0 ~ \ldots ~ \aaa_{\Khat} ] \in \cset{NM}{(\Khat+1)}$ and $ \aaa_k \triangleq \cc^\conj(\nuhat_k) \otimes \bb(\tauhat_k)$. Then, the pilot-only channel estimate can be refined via reconstruction from the target parameter estimates as \vspace{-0.1in}
\begin{align} \label{eq_hhat_stage1}
    \HHhat = \sum_{k=0}^{\Khat}   \alphahat_{k} \,  \bb(\tauhat_{k}) \cc^\trp(\nuhat_{k}) \, .
\end{align}

\subsection{Stage 2: Data Demodulation}
We propose to employ the LMMSE estimator for data demodulation from \eqref{eq_YY} using the channel estimate either from \eqref{eq_hhat_stage1} at Stage~1 (initial phase) or from \eqref{eq_rf}, \eqref{eq_mf} or \eqref{eq_lmmse_channel} at Stage~3 (during iterations) \cite[Eq.~(2.41)]{fazel2008multi}, \cite[p.~389]{kay1993fundamentals}, \cite{holisticISAC_2024}:
\begin{align} \label{eq_lmmse_data}
        \XXdhat =  \big( \YYdhat \odot \HHdhat^\conj \big) \oslash \big( \abs{\HHdhat}^2 + \snrx^{-1}  \big) \,,  
\end{align}
where $\snrx \triangleq \Eee\{ \abss{\xnm}^2 \} / \sigma^2 = 1 / \sigma^2 $.

\subsection{Stage 3: Channel Estimation via Pilots and Demodulated Data}
Now that we have obtained data estimates in \eqref{eq_lmmse_data} at Stage~2, we can now refine $\HHdhat$ via the following three estimators commonly employed in the OFDM ISAC literature (while $\HHphat$ remains fixed to its value in \eqref{eq_hhhat_p}).

\subsubsection{Reciprocal Filtering (RF)}
\ac{RF} performs element-wise division of received symbols by transmit symbols \cite{RadCom_Proc_IEEE_2011,OFDM_Radar_Phd_2014,OFDM_Radar_Corr_TAES_2020,5G_NR_JRC_analysis_JSAC_2022,PRS_ISAC_5G_TVT_2022}:
 \begin{align}\label{eq_rf}
        \HHdhat= \YYdhat \oslash \XXdhat  \,.
\end{align}

\subsubsection{Matched Filtering (MF)}\label{sec_mf}
\ac{MF} performs matched filtering of received symbols with transmit symbols \cite{OFDM_Radar_Corr_TAES_2020,reciprocalFilter_OFDM_2023}:
\begin{align} \label{eq_mf}
        \HHdhat= \YYdhat \odot \XXdhat^\conj \,.
\end{align}

\subsubsection{LMMSE Estimator}\label{sec_mf}
The LMMSE estimator relies on the assumption that delays and Dopplers in the ISAC channel $\HH$ in \eqref{eq_hh} are uniformly distributed in their respective unambiguous detection intervals, and is given by \cite{holisticISAC_2024}
\begin{align} \label{eq_lmmse_channel}
        \HHdhat =  \big( \YYdhat \odot \XXdhat^\conj \big) \oslash \big( \abs{\XXdhat}^2 + \snrh^{-1}  \big) \,,  
\end{align}
where $\snrh \triangleq \sum_{k=0}^K \abs{\alpha_k}^2 /\sigma^2$.

The resulting $\HHdhat$ from \eqref{eq_rf}, \eqref{eq_mf} or \eqref{eq_lmmse_channel} is fed back to Stage~2 to refine data estimates in \eqref{eq_lmmse_data}. The iterations between Stage~2 and Stage~3 continue until a termination criterion is satisfied (e.g., maximum number of iterations).

\subsection{Stage 4: Target Detection and Estimation from Unstructured ISAC Channels}
At the final stage, $\HHdhat$ from Stage~3 and $\HHphat$ in \eqref{eq_hhhat_p} are combined to construct the final ISAC channel estimate $\HHhat$. For target detection/estimation from $\HHhat$, we apply the procedure in Sec.~\ref{sec_det_est} to obtain the ultimate delay-Doppler detections.

\section{Numerical Results}\label{sec_num}
\subsection{Simulation Setup}
\subsubsection{Scenario and Parameters}
We evaluate the performance of the proposed bistatic ISAC receiver in Sec.~\ref{sec_rec_design} using the default simulation parameters in Table~\ref{tab_SimParam}. The ISAC-TX beamforming vector in \eqref{eq_ynm} is set to point towards $10^\circ$, i.e., $\fftx = \atx^\conj(10^\circ)$. For the unified ISAC channel in \eqref{eq_hh_vec}, we assume the presence of $K=2$ targets along with a \ac{LOS} path, where 
the channel gains are generated as
\vspace{-0.1in}
\begin{align}
    \alpha_0 = \frac{\lambda}{4 \pi d_0} \,, \, \alpha_k = \frac{\lambda \sqrt{\sigmarcsk}}{ (4 \pi)^{3/2} d_{k,1} d_{k,2}} \, , k > 0 \,,
    \vspace{-0.2in}
\end{align}
with $d_0 = \norm{\ptx - \prx}$, $d_{k,1} = \norm{\ptx - \ptk}$, $d_{k,2} = \norm{\ptk - \prx}$ and $\sigmarcsk$ denoting the bistatic \ac{RCS} of the $\thn{k}$ target. Here, $\ptx = [0, 0] \, \rm{m}$, $\prx = [50, 0] \, \rm{m}$, $\pp_1 = [56.9, 10] \, \rm{m}$ and $\pp_2 = [79.4, 7] \, \rm{m}$ denote the locations of ISAC-TX, ISAC-RX, Target-$1$ and Target-$2$, respectively. ISAC-TX and ISAC-RX are stationary, while the target velocities are set to $\vv_1 = [1.4, -2.2] \, \rm{m/s}$ and $\vv_2 = [2.2, -13.7] \, \rm{m/s}$. Moreover, we set the false alarm probability of the CFAR detector implemented on $\hhatdd$ in \eqref{eq_hhatdd} to $\pfa = 10^{-4}$. Unless otherwise stated, we employ QPSK modulation for $\XXd$, set the RCSs to $\sigma_{{\rm{rcs}},1} = 4.9 \, \rm{dBsm}$ and $\sigma_{{\rm{rcs}},2} = 1.5 \, \rm{dBsm}$, and the pilot percentage to $\prat = 5 \%$. The pilots are randomly placed in the frequency-time grid and kept fixed throughout the simulations. 

\subsubsection{Benchmarks}
For benchmarking purposes, we assess the ISAC performance of the following algorithms:
\begin{itemize}
    \item \textbf{Proposed -- Data-Aided:} The proposed multi-stage data-aided iterative sensing algorithm in Sec.~\ref{sec_rec_design}, with three different flavors (RF, MF or LMMSE) in Stage~3.
    \item \textbf{Benchmark 1 -- Pilot-Only:} Pilot-only target detection/estimation, as described in Sec.~\ref{sec_ce_pilot1} and Sec.~\ref{sec_det_est}. This sets a \textit{lower bound} on the performance of the proposed algorithm.
    \item \textbf{Benchmark 2 -- Genie:} Assuming perfect knowledge of data symbols, i.e., $\XXdhat = \XXd$, we apply \eqref{eq_hhhat_p} at pilot locations and LMMSE in \eqref{eq_lmmse_channel} at data locations in Stage~3, and execute Stage~4. This serves as an \textit{upper bound} on the performance of the proposed algorithm.    
\end{itemize}

\subsubsection{Metrics}
We use the following metrics for evaluation.
\begin{itemize}
    \item \textbf{Sensing Metric:} We generate $500$ independent Monte Carlo realizations of the noise matrix $\ZZ$ in \eqref{eq_YY} and calculate the empirical probability of detection ($\pd$) as our sensing metric.
    \item \textbf{Communication Metric:} We adopt the achievable rate as the communication metric, calculated via the \ac{MI} $\Iind(\XX;\YY \, | \, \HH)$ between $\XX$ and $\YY$ in \eqref{eq_YY} for a given constellation of data symbols $\XXd$ \cite[Alg.~3]{holisticISAC_2024}. The achievable rate is given by $\Iind(\XX;\YY \, | \, \HH) (100-\prat)/100$ for all the considered algorithms.\footnote{\label{fn_ch_est}In our analysis of achievable rates and capacity, we deliberately exclude the impact of imperfect channel estimation to focus solely on the effects of reduced data payload as a consequence of increasing pilot ratio, which allows isolated investigation of such effects on communication metrics, independent of channel estimation quality. The impact of channel estimation quality has already been extensively covered in OFDM communications literature \cite{OFDM_Pilot_Giannakis_2004}.}\footnote{We note that the considered algorithms differ only in their sensing performances based on how they use pilot and/or data for channel and target parameter estimation, while their (shared) communication performance is measured by the percentage of data payload (i.e., $100-\prat$) and the \ac{MI} $\Iind(\XX;\YY \, | \, \HH)$ under the assumption of perfect knowledge of $\HH$.}
\end{itemize}

{\setlength{\abovecaptionskip}{11 pt}

\begin{table}
    \caption{Simulation Parameters}
    \vspace{-0.1in}
    \label{tab_SimParam}
    \centering
    \begin{tabular}{llr}
         \textbf{Parameter} & \textbf{Description} & \textbf{Value}  \\ \hline \hline

         $\fc$ & Carrier frequency & $28 \, \rm{GHz}$ \\
         $\deltaf$ & Subcarrier spacing & $120 \, \rm{kHz}$ \\
         $N$ & \# subcarriers & $400$ \\
         $N\deltaf$ & Bandwidth & $48 \, \rm{MHz}$ \\
         $M$ & \# symbols & $60$ \\
         $\Pt$ & Transmit power  & $20 \, \rm{dBm}$ \\
         $\Ntx$ & ULA size at ISAC-TX  & $8$   \\
         $N_0$ & Noise PSD & $-174 \, \rm{dBm/Hz}$   \\
         $N_F$ & Noise figure & $8 \, \rm{dB}$   \\
         \hline
    \end{tabular}
\end{table}}

\subsection{Impact of Target RCS}
\begin{figure}[t]
	\centering
    \vspace{-0.1in}
	\includegraphics[width=1\linewidth]{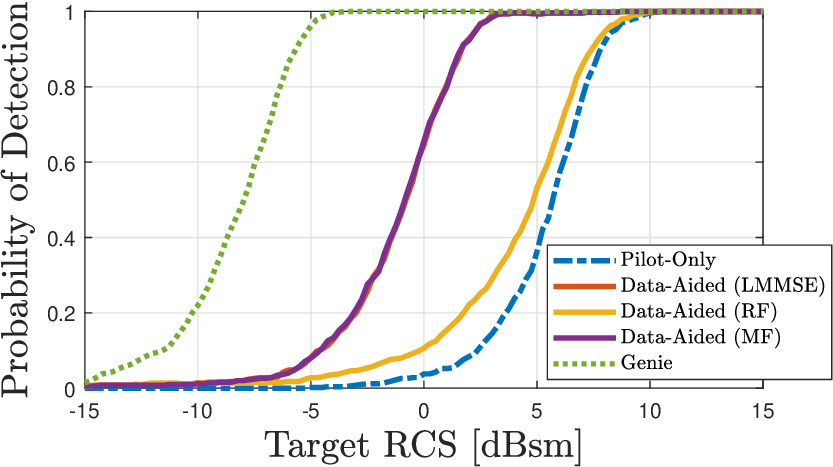}
	\vspace{-0.3in}
	\caption{Probability of detection with respect to target RCS for $\prat = 5 \%$.} 
	\label{fig_RCS}
	\vspace{-0.1in}
\end{figure}

We first investigate the detection performance of various bistatic ISAC receivers, with the goal of evaluating how the strong target (Target-$1$) masks the weak one (Target-$2$) due to increased sidelobe levels resulting from a lack or imperfect knowledge of $\XXd$. To that end, Fig.~\ref{fig_RCS} shows $\pd$ of Target-$2$ as a function of its \ac{RCS}. As expected, the genie-aided receiver with perfect knowledge of $\XXd$ achieves the highest $\pd$ performance, while the pilot-only scheme yields the poorest one. The proposed data-aided sensing receiver with LMMSE and MF provides significant gains over the pilot-only scheme (up to $7 \, \rm{dBsm}$ improvement), closing half the gap to the genie-aided case (that can reach up to $14 \, \rm{dBsm}$), which highlights its effectiveness in recovering performance losses due to unknown $\XXd$.

\subsection{Impact of Pilot Percentage}
\begin{figure}[t]
	\centering
	\includegraphics[width=0.95\linewidth]{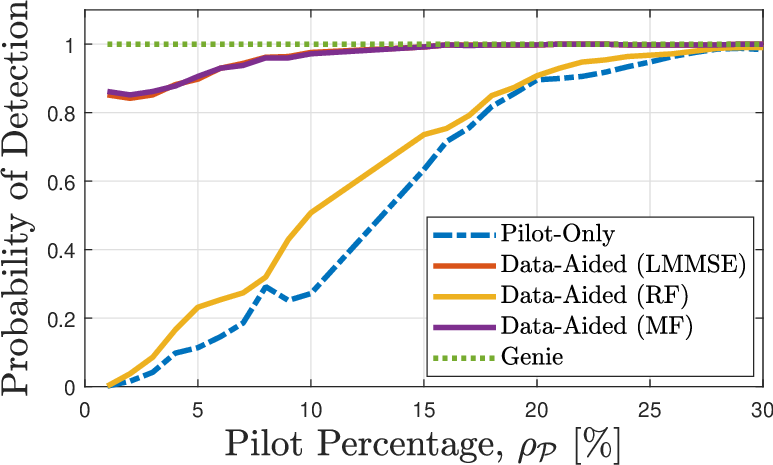}
	\vspace{-0.1in}
	\caption{Probability of detection with respect to pilot percentage for $\sigma_{{\rm{rcs}},2} = 2 \, \rm{dBsm}$.} 
	\label{fig_pilot_perc}
	\vspace{-0.1in}
\end{figure}

\begin{figure}[t]
	\centering
	\includegraphics[width=0.95\linewidth]{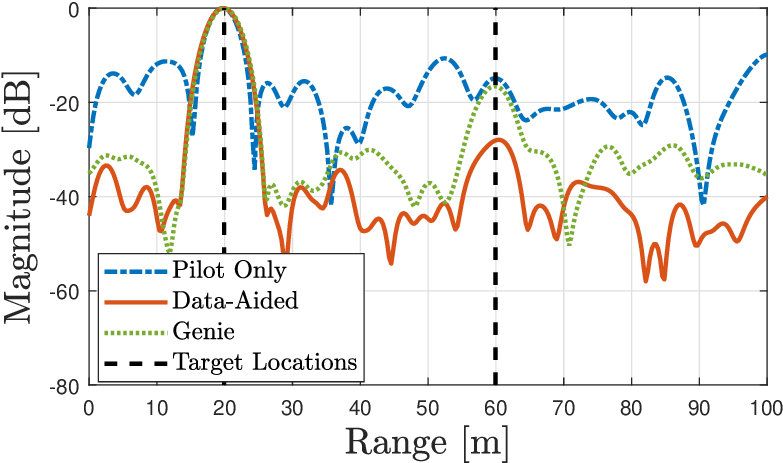}
	\vspace{-0.1in}
	\caption{Range profiles obtained by pilot-only, data-aided (LMMSE) and full-data (genie-aided) schemes for $\sigma_{{\rm{rcs}},2} = 2 \, \rm{dBsm}$ and $\prat = 2 \%$. The range values represent the differential ranges with respect to the \ac{LOS} path.} 
	\label{fig_range_profile}
	\vspace{-0.1in}
\end{figure}

We now evaluate the impact of pilot percentage $\prat$ on $\pd$ of Target-$2$ for fixed RCS $\sigma_{{\rm{rcs}},2} = 2 \, \rm{dBsm}$, reported in Fig.~\ref{fig_pilot_perc}, using the same setup as in Fig.~\ref{fig_RCS}. It is observed that the performance of the proposed data-aided sensing algorithm converges quickly to that achievable under perfect knowledge of data symbols (i.e., genie) as $\prat$ increases, substantially outperforming the pilot-only benchmark and closing almost all the gap to the genie-aided one when $\prat$ exceeds $10 \%$. Similar to Fig.~\ref{fig_RCS}, RF performs poorly compared to LMMSE and MF due to its vulnerability to low SNR conditions (severe AWGN and/or demodulation noise) \cite{holisticISAC_2024}. In addition, to illustrate how the different schemes affect $\pd$, we provide an example of range profiles in Fig.~\ref{fig_range_profile} obtained via \eqref{eq_hhatdd}. We observe that the pilot-only scheme results in extremely high sidelobe levels due to \textit{(i)} low SNR and \textit{(ii)} holes in the frequency-time spectrum \cite{MIMO_OFDM_radar_TAES_2020,ISAC_OFDM_Pilots}, leading to masking of the weak target. On the other hand, the proposed data-aided sensing algorithm can significantly reduce sidelobe levels and successfully recover the weak target, bridging the gap to the genie-aided scheme.

\subsection{Bistatic ISAC Trade-offs}
\begin{figure}[t]
	\centering
    \vspace{-0.1in}
	\includegraphics[width=0.95\linewidth]{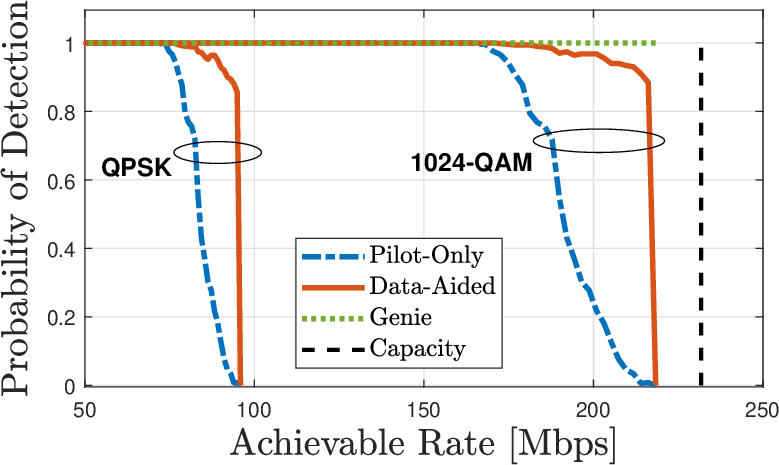}
	\vspace{-0.1in}
	\caption{Bistatic ISAC trade-off curves with QPSK and $1024$-QAM modulation, obtained by sweeping $\prat$ over $[0, 100] \%$. The data-aided receiver employs the LMMSE estimator at Stage 3.} 
	\label{fig_tradeoff}
	\vspace{-0.1in}
\end{figure}

To explore the trade-offs between $\pd$ and achievable rate in bistatic ISAC, we sweep $\prat$ from $0 \%$ to $100 \%$ under QPSK and $1024$-QAM modulation for $\XXd$. The resulting trade-off curves are reported in Fig~\ref{fig_tradeoff}, where the horizontal genie-aided $\pd$ line and the vertical capacity line delineate the boundary of the ISAC trade-off region. Using higher-order modulations improves the achievable rate, as expected, and the proposed data-aided scheme can significantly enhance ISAC trade-offs, extending the trade-off curves closer to the optimal boundary compared to the pilot-only approach. 

\subsection{Performance Over Iterations}
\begin{figure}[t]
	\centering
	\includegraphics[width=0.95\linewidth]{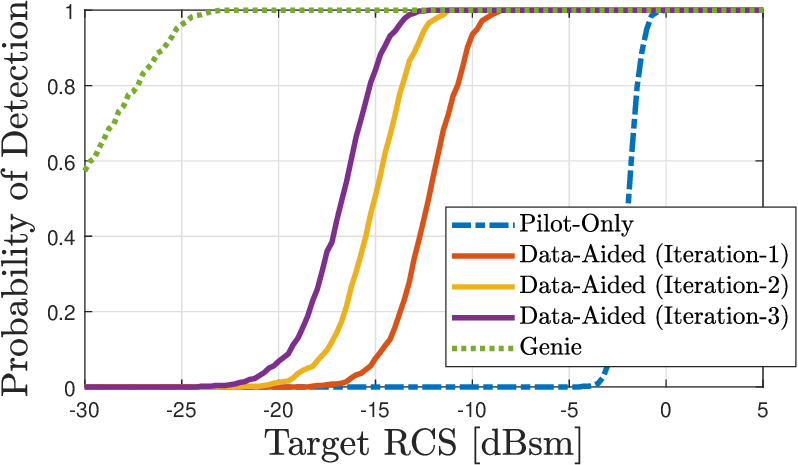}
	\vspace{-0.1in}
	\caption{Evolution of probability of detection achieved by LMMSE over iterations of the data-aided sensing algorithm for $\prat = 5 \%$.} 
	\label{fig_iter}
	\vspace{-0.2in}
\end{figure}

We finally investigate how the performance of the proposed data-aided sensing scheme with LMMSE evolves over iterations. Fig.~\ref{fig_iter} shows $\pd$ of Target-$2$ for $\prat = 5 \%$ and $\Pt = 40 \, \rm{dBm}$ relative to its \ac{RCS}. We observe improved detection performance over iterations (up to $5 \, \rm{dBsm}$ improvement in two iterations), showing the promising performance of the iterative data-aided sensing approach. 

\section{Concluding Remarks}
We have introduced a data-aided sensing approach for bistatic OFDM ISAC systems, showcasing its significant potential to enhance ISAC performance. In particular, the proposed algorithm can substantially close the vast gap in detection performance between pilot-only schemes (relying solely on pilots for sensing) and genie-aided ones (exploiting both pilots and perfect knowledge of data for sensing), effectively addressing the challenge of unknown data payload in bistatic ISAC systems. Future research will focus on developing ISAC signal design strategies (i.e., TX beamforming and constellation design) to optimize bistatic ISAC trade-offs.

\bibliographystyle{IEEEtran}
\bibliography{IEEEabrv,Sub/main}

\end{document}